\documentclass[twocolumn,prd,preprintnumbers,amsmath,amssymb]{revtex4}

\usepackage{graphicx}
\usepackage{dcolumn}
\usepackage{bm}
\usepackage{epsfig}
\usepackage{amssymb}
\usepackage{amsmath}

 \def\tskip{\setlength{\tskip}{5pt}}
\def\colwidth{\setlength{\colwidth}{3.5in}}

\def\prd{Phys. Rev. D}
\def\pr{Phys. Rep.}
\def\prl{Phys. Rev. Lett.}

\def\apj{Astrophys. J.}
\def\apjl{Astrophys. J. Lett.}

\def\mnras{Mon. Not. Roy. Astron. Soc.}

\newcommand{\lsim}{\mathrel{\hbox{\rlap{\lower.55ex\hbox{$\sim$}} \kern-.3em \raise.4ex \hbox{$<$}}}}
\newcommand{\gsim}{\mathrel{\hbox{\rlap{\lower.55ex\hbox{$\sim$}} \kern-.3em \raise.4ex \hbox{$>$}}}}
\newcommand{\beq}{\begin{equation}}
\newcommand{\eeq}{\end{equation}}
\newcommand{\beqa}{\begin{eqnarray}}
\newcommand{\eeqa}{\end{eqnarray}}

\begin{document}

\title{Relic gravitational waves: latest revisions and preparations for new data}

\author{Wen Zhao$^{1,2}$ and L. P. Grishchuk$^{1,3}$}
\affiliation{
$^{1}$School of Physics and Astronomy, Cardiff University,
Cardiff, CF24 3AA, United Kingdom\\
$^{2}$Wales Institute of Mathematical and Computational Sciences,
Swansea, SA2 8PP, United Kingdom\\
$^{3}$Sternberg Astronomical Institute, Moscow State University, Moscow,
119899, Russia}

\date{\today}

\begin{abstract}

The forthcoming release of data from the Planck mission, and
possibly from the next round of Wilkinson Microwave Anisotropy
Probe (WMAP) observations, make it necessary to revise the
evaluations of relic gravitational waves in the existing data and,
at the same time, to refine the assumptions and data analysis
techniques in preparation for the arrival of new data. We
reconsider with the help of the commonly used CosmoMC numerical
package the previously found indications of relic gravitational
waves in the 7-year (WMAP7) data. The CosmoMC approach reduces the
confidence of these indications from approximately 2$\sigma$ level
to approximately 1$\sigma$ level, but the indications do not
disappear altogether. We critically analyze the assumptions that
are currently used in the cosmic microwave background (CMB) data
analyses and outline the strategy that should help avoid the
oversight of relic gravitational waves in the future CMB data. In
particular, it is important to keep away from the unwarranted
assumptions about density perturbations. The prospects of
confident detection of relic gravitational waves by the Planck
satellite have worsened, but they are still good. It appears that
more effort will be required in order to mitigate the foreground
contamination.

\end{abstract}


\maketitle


\section{Introduction \label{section-1}}

Cosmology relies on extrapolations. It is natural to extrapolate
the picture of approximate homogeneity and isotropy seen in our
patch of the Universe to other places and times. Having assumed
the usual framework of a homogeneous and isotropic hot big bang,
we do not encounter any real contradictions or paradoxes requiring
drastic solutions. But we do encounter an initial cosmological
singularity \cite{hawk}. The singularity is not an ultimate
answer, it is only a sign of limited applicability of the
currently available theories. The limit of applicability is
probably set by the Planck parameters. It seems logical to suggest
that our Universe came into being as a configuration with a
Planckian size and a Planckian energy density, and with a total
energy, including gravity, equal to zero (see \cite{birthpapers}
and references therein).

It is here where a real problem arises. The newly created
classical configuration cannot reach the averaged energy density
and size of the presently observed Universe, unless the
configuration experienced a kind of primordial kick. During the
kick the size of the newly born Universe (for simplicity, one can
think of the curvature radius of a closed universe) should
increase by many orders of magnitude without significant changes
in the energy density of the whatever substance that was there.
Such an expansion cannot be driven by normal types of matter which
we presently know. The kick should be driven by something more
exotic: a lucky version of a scalar field (inflaton) \cite{guth}
or something like a conformal anomaly \cite{starob} or something
from the ``theory of everything".

The strong variable gravitational field of the very early Universe
inevitably generates gravitational waves and, under certain extra
conditions, density perturbations and rotational perturbations
\cite{grishchuk1974}. The generating mechanism is the
superadiabatic (parametric) amplification of the zero-point
quantum oscillations of the respective degrees of freedom, or, in
more technical terms, the Schrodinger evolution of the initial
vacuum (ground) state of the corresponding time-dependent
Hamiltonian into a strongly squeezed vacuum (multiparticle)
quantum state \cite{grishchuk1974}. The generated gravitational
waves and density perturbations leave observable imprints in the
temperature and polarization anisotropies of the cosmic microwave
background radiation. By studying the CMB correlation functions
$TT$, $TE$, $EE$, $BB$ we can learn about the birth of the
Universe and the initial stage of cosmological expansion.

The simplest assumption about the initial kick is that its entire
duration can be described by the scale factor with one fixed
power-law dependence \cite{grishchuk1974}:
\begin{equation}
\label{inscf}
a(\eta) = l_o|\eta|^{1+ \beta},
\end{equation}
where $l_o$ and $\beta$ are constants, $\beta < -1$. Then, the generated
primordial spectra of metric perturbations describing gravitational
waves (t) and density perturbations (s) (for more explanations,
see \cite{zbg}) have the universal power-law forms
\begin{eqnarray}
\label{PsPt}
P_{t}(k)=A_t \left(\frac{k}{k_0}\right)^{n_t},~~~
P_{s}(k)=A_s \left(\frac{k}{k_0}\right)^{n_s-1},
\end{eqnarray}
where we will be using $k_0=0.002$Mpc$^{-1}$. According to the theory of
quantum-mechanical generation of cosmological perturbations
\cite{grishchuk1974} the spectral
indices are approximately equal, $n_s-1 = n_t = 2(\beta+2)$, and the
amplitudes $(A_t)^{1/2}$ and $(A_s)^{1/2}$ are of the order of magnitude
of the ratio $H_i/H_{\rm Pl}$, where $H_i\sim c/l_o$ is the
characteristic value of the Hubble parameter during the kick.

The presently available CMB and large-scale structure observations
cover a huge interval of scales, ranging from wave numbers $k
\approx 0.0002$Mpc$^{-1}$ and up to $k \approx 0.2$Mpc$^{-1}$.
There is no any particularly fundamental reason why a single
power-law evolution (\ref{inscf}) should be valid during a long
interval of time and, consequently, exact power-law spectra
(\ref{PsPt}) valid in the interval of wavelengths where the ends
of the spectrum differ from each other by a factor $10^3$. We do
not know such spectra in physics and astrophysics. In cosmology,
one can also expect some deviations from strict power-laws
(\ref{PsPt}) \cite{grishchuk1991}, \cite{kosowsky1995}.

If the continuous power spectrum can be approximated by two
power-law pieces, one can write \beqa \label{break1}
P_s(k)&=&A_{s}^{(1)}\left(\frac{k}{k_0}\right)^{n_s^{(1)}-1},~~~~~~k<k_1, \\
\label{break2}
P_s(k)&=&A_{s}^{(2)}\left(\frac{k}{k_0}\right)^{n_s^{(2)}-1},~~~~~~k>k_1,
\eeqa
where $A_{s}^{(2)}=A_{s}^{(1)}\left(\frac{k_1}{k_0}
\right)^{n_s^{(1)}-n_s^{(2)}}$, and
\beqa
\label{break3}
P_t(k)&=&A_{t}^{(1)}\left(\frac{k}{k_0}\right)^{n_t^{(1)}},~~~~~~k<k_1, \\
\label{break4}
P_t(k)&=&A_{t}^{(2)}\left(\frac{k}{k_0}\right)^{n_t^{(2)}},~~~~~~k>k_1,
\eeqa
where $A_{t}^{(2)}=A_{t}^{(1)}\left(\frac{k_1}{k_0}
\right)^{n_t^{(1)}-n_t^{(2)}}$.

Obviously, we return to Eq.(\ref{PsPt}) if the spectral indices in the
two pieces are equal to each other.

Alternatively, one can postulate a simple law of ``running" of the spectral
index,
\begin{eqnarray}
\label{runns}
n_{s}(k) = n_{s}(k_0) +\alpha_s \ln(k/k_0),
\end{eqnarray}
where $\alpha_s$ is a constant. The spectral index $n_{s}(k)-1$ for density
perturbations is defined as $n_{s}(k)-1 = [d~\ln P_{s}(k)/d~\ln k]$, and
similar definition is used for $n_t(k)$. Then the power spectrum for density
perturbations takes the form
\begin{eqnarray}
\label{runPs}
P_{s}(k) = A_{s}(k_0) \left(\frac{k}{k_0}\right)^{n_{s}(k_0) -1 +
\frac{1}{2}\alpha_s \ln(k/k_0)},
\end{eqnarray}
and a similar formula holds for gravitational waves. In our calculations
below we will be testing both options, (\ref{break1})-(\ref{break2}) and
(\ref{runns}), of spectral deviations.

Following the (quite unfortunate) tradition, we are characterizing
the amount of relic gravitational waves (GW) by ratios involving
density perturbations (DP), rather than by the gravitational wave
amplitudes directly. One measure is the quadrupole ratio
\begin{eqnarray}
\label{defineR}
R \equiv \frac{C_{\ell=2}^{TT}({\rm gw})}{C_{\ell=2}^{TT}({\rm dp})},
\end{eqnarray}
another -- the so-called tensor-to-scalar ratio
\begin{eqnarray}
\label{definer}
r \equiv \frac{A_t(k_0)}{A_s(k_0)}.
\end{eqnarray}
By definition, $r$ and $R$ cannot be negative. Some elements of numerical
software are better adjusted to one quantity than to another. We will be
using both of them remembering that $r\approx 2R$ for a quite wide class of
models~\cite{zbg}.

Contrary to the theory of quantum-mechanical generation of cosmological
perturbations, the inflationary theory claims that the power spectrum of
density perturbations should be many orders of magnitude larger than
the power spectrum of gravitational waves. The inflationary theory attempts
to apply to density perturbations the same mechanism of superadiabatic
(parametric) amplification that was originally worked out for gravitational
waves \cite{grishchuk1974}. But the so-called classic result of
inflationary theory states that the $A_s$ should be
arbitrarily large in the limit of de Sitter inflation (i.e. when the
perturbations are generated with spectral indices $n_s =1$ and $n_t = 0$):
$A_s \approx A_t/ \epsilon$, where $\epsilon \equiv - {\dot H}/{H^2}$ is
zero for de Sitter expansion with any Hubble parameter $H$ (i.e. for any
``energy scale of inflation"). The existence and properties of primordial
density perturbations are consequences of the quantum-mechanical generating
mechanism, and not of the inflationary theory. The contribution of
inflationary theory to this subject is the (incorrect) prediction of
divergent $A_s$, that is, the (incorrect) prediction of amplitudes that can be
arbitrarily larger than $H_i/H_{\rm Pl}$. For inflationary considerations
about density perturbations, see for example \cite{mukh} and recent books
on cosmology.

Inflationists rarely suffer from the problem of ``too little", they usually
have the fine-tuning trouble because they generate ``too much". They have
started from the threat to ``overclose the Universe" and they continue this
trend till now. To make the wrong theory look ``consistent", inflationists
convert their prediction of arbitrarily large $A_s$ into the prediction of
arbitrarily small $A_t$. As a result, the most advanced string-based
inflationary theories predict the negligibly small amounts of gravitational
waves, $r \approx 10^{-24}$ or less. In other words, inflationists attempt to
pay by gravitational waves for their incorrect treatment of density
perturbations. (For a more detailed criticism of inflationary theory, see
the last paper in \cite{grishchuk1974}.) Obviously, we do not accept the
inflationary theory. We allow the use of the inflationary
``consistency relation" $r = -8n_t$ only in Sec.\ref{section-2} where we
repeat the derivations of the WMAP Collaboration \cite{komatsu},
\cite{larson} who used this relation.

The previous search \cite{zbg} for relic gravitational waves in
the WMAP data was based on the simplified likelihood function
which, among other things, neglects the data and noise
correlations at the lowest multipoles. The approaching release of
data from the Planck mission \cite{planck} and, possibly, from the
next round of WMAP observations, makes it necessary to test by
other numerical techniques the previously found indications of
relic gravitational waves. The WMAP team and other groups
routinely employ the CosmoMC software \cite{cosmomc}. The data
analysis constructions in \cite{zbg} are based on the genuine
Wishart probability density function, whereas the WMAP likelihood
function, as a part of CosmoMC package, is based on semi-Gaussian
approximations to this distribution. The new analysis with CosmoMC
is certainly different from the previous one, but we are reluctant
to say that it is necessarily more accurate from the physical
point of view. However, we realize that the CosmoMC code is the
approach that most of the groups are pursuing now and will be
using in the future with new data. Therefore, it is important to
revise the previous evaluations of relic gravitational waves with
the help of CosmoMC, and to make revised forecasts for Planck
observations. This is the main purpose of this paper.

In Sec.\ref{section-2} we rederive the results of WMAP
Collaboration \cite{komatsu}, \cite{larson} using their
assumptions and methods. In Sec.\ref{sec3} we apply the CosmoMC
package to repeat the previously performed search for relic GW at
lower-$\ell$ multipoles \cite{zbg}. We show that the indications
of GW diminish but do not disappear altogether. In Sec.\ref{sec4}
we demonstrate that the conclusions about $n_s$ and $r$ depend on
the interval of multipoles utilized in the data analysis. It is
unwise to postulate constant spectral indices in a huge interval
of accessible wavelengths and multipoles. In Sec.\ref{sec5} we
explore a hypothesis of primordial spectra consisting of two
power-law pieces. It is shown that the hypothesis is consistent
with the data and with the results discussed in previous Sections.
Sec.\ref{sec6} is devoted to refined forecasts for the Planck
mission. The prospects of confident detection of relic
gravitational waves by Planck are still reasonable.

\section{Repeating the WMAP7 analysis \label{section-2}}

First of all, with the help of CosmoMC, we repeat the WMAP7
calculations. We want to be sure that we obtain the same results
when we make the same assumptions. Obviously, we include
gravitational waves from the very beginning, we do not consider
the ``minimal" model of WMAP \cite{komatsu} where gravitational
waves are voluntarily excluded from the outset. In Sec.
\ref{const}, the spectral indices are assumed strictly constant in
the entire interval of relevant wavelengths, that is, together
with the WMAP team, we adopt Eq.(\ref{PsPt}). In Sec. \ref{run},
we allow for the ``running" of $n_s$, that is, together with WMAP,
we use Eq.(\ref{runns}). Moreover, in these two subsections, and
only there, instead of the correct relation $n_t = n_s -1$, we use
the incorrect ``consistency relation" $n_t = -r/8$ for reducing
the number of unknown parameters. This is done exclusively because
the WMAP team has done this, and we want to rederive their
results. Since $r$ cannot be negative, the inflationary theory
does not allow positive $n_t$.

\subsection{Constant spectral indices \label{const}}

We repeat the WMAP team analysis of the 7-year CMB data with the help of the
CosmoMC sampler. The space of free parameters subject to evaluation consists
of four background parameters ($\Omega_bh^2$, $\Omega_{c}h^2$, $\tau$,
$\theta$) and three perturbation parameters ($\ln (10^{10}A_s)$, $n_s$,
$r$). These are the standard CosmoMC parameters for a flat $\Lambda$CDM
cosmology.  As mentioned, we also adopt Eq.(\ref{PsPt}) and atop of that,
$n_t = -r/8$. We call this set of assumptions the {\it Case I}.

The analysis takes into account the observed CMB data for all four angular
power spectra: $TT$, $TE$, $EE$ and $BB$ \cite{lambda}. The range of the
used multipoles is  $2\le\ell\le\ell_{\rm max}$, where
$\ell_{\rm max}=1200$ for $TT$ spectrum, $800$ for $TE$ spectrum,
and $23$ for $EE$ and $BB$ spectra. The maximum likelihood values and
marginalized distributions are found for all seven free parameters, but we
are showing the results only for $n_s$ and $r$ because we are mostly
interested in them.

The results of the CosmoMC calculations for the {\it Case I} are
summarized in the upper row of Table \ref{table1} and in
Fig.~\ref{figure1}. The seven-dimensional (7d) ML values of $n_s$
and $r$ are shown in the column ``Maximum likelihood". The
one-dimensional (1d) marginalized distributions for $n_s$ and $r$
are shown in Fig.~\ref{figure1} by olive curves, marked also by
symbol (1). (The maximum of 1d likelihoods is always normalized to
1.) The obtained results are very close to the WMAP findings
\cite{komatsu}. We see that the assumptions of the {\it Case I}
lead to the conclusion that the primordial spectrum is red, $n_s
<1$. Although the uncertainties are still large, the $n_s <1$ is
the preferred outcome of the WMAP analysis \cite{komatsu} and of
our repetition here under the same assumptions. The $n_s <1$ would
mean that the function $P_{s}(k)$ is infra-red divergent at very
small wavenumbers $k$. The maximum likelihood (ML) values of $r$,
both, for the 7d and 1d posterior distributions, are very close to
zero. The $95\%$ confidence limit $r \approx 0.38$ is usually
interpreted as the upper limit for the possible amount of relic
gravitational waves.

\subsection{Running spectral index \label{run}}

Here, together with the WMAP Collaboration, we increase the number of free
parameters from seven to eight by adding the parameter $\alpha_s$ and adopting
Eq.(\ref{runns}). We call this set of assumptions the {\it Case II}. Other
parameters, observational data, and CosmoMC numerical techniques are exactly
the same as in the {\it Case I}. The parameter $\alpha_s$ does not have to be
a nonzero number. It is the maximum likelihood analysis of the data that
determines its value.

The results of the analysis for the {\it Case II} are shown in the
lower row of Table \ref{table1} and in Fig. \ref{figure1}. The
eight-dimensional (8d) ML values of $n_s(k_0)$, $\alpha_s$ and $r$
are shown in the column ``Maximum likelihood". The posterior 1d
distributions of $n_s(k_0)$ and $r$ are plotted by magenta curves,
marked also by symbol (2).

Together with the WMAP team, we see that the primordial spectrum in the
{\it Case II} becomes blue at long wavelengths, that is, the 8d and
1d ML values of $n_{s}(k_0)$ are larger than 1. The power spectrum
$P_{s}(k)$ is no longer divergent at very small $k$. Since the ML
$\alpha_s$ is negative, the spectrum gradually turns over from the blue at
long wavelengths to red, $n_{s}(k)<1$, at shorter wavelengths, see also
Eq.(\ref{runns}). There is a considerable decrease of $n_s(k)$ on the way
from the long-wavelength end of the spectrum at $k \approx 0.0002$Mpc$^{-1}$
to the short-wavelength end of the spectrum at $k \approx 0.2$Mpc$^{-1}$.
(We restrain from a discussion of disastrous consequences that the
values $n_{s}(k) \ge 1$ bring forth to the inflationary theory.)

In comparison with the {\it Case I}, the 1d distribution of $r$
becomes significantly broader, and it is almost flat at small values of
$r$, where $r\in(0,~0.12)$. The 8d ML value of $r$ is $r=0.06$. This value is
7.5 times larger than that in the {\it Case I}, but one should take this
number with caution. When the distribution is almost flat, the ML point can
be accidental. It is clear from the shape of the magenta curve in
Fig. \ref{figure1} that the assumptions of the {\it Case II} are
consistent with the hypothesis of no gravitational waves, $r=0$.
Nevertheless, the probability of large values of $r$ has increased, as
compared with the {\it Case I}. The $95\%$ C.L. has risen up to
$r \approx 0.43$.

\begin{table*}
\caption{Results for $n_s$, $\alpha_s$ and $r$ in Case I and Case II}
\begin{center}
\label{table1}
\begin{tabular}{|c|c|c|c|c|c|c|c|}
  \hline
  \multicolumn{1}{|c|}{  } & \multicolumn{3}{c|}{Maximum likelihood}& \multicolumn{3}{c|}{1-d likelihood}  \\
         \hline
 & $n_s$& $\alpha_s$ & $r$& $n_s$ &$\alpha_s$& $r$~($95\%$C.L.)\\
          \hline
 Case I & $0.967$ &$\cdot\cdot\cdot$& $0.008$& $0.991^{+0.021}_{-0.020}$ &$\cdot\cdot\cdot$& $r<0.379$ \\
           \hline
 Case II & $1.061$ &-0.043& $0.060$& $1.065^{+0.058}_{-0.058}$ &$-0.039^{+0.029}_{-0.026}$& $r<0.430$\\
  \hline
\end{tabular}
\end{center}
\end{table*}

\section{Evaluations of relic gravitational waves from lower-$\ell$ CMB data
\label{sec3}}

The purpose of this Section is to revise by the CosmoMC code the
previously found indications of relic gravitational waves
\cite{zbg} in the WMAP7 data. It was stressed many times
\cite{zbg}, \cite{kam2005} that the relic GW compete with density
perturbations only at relatively small multipoles $\ell$. It is
dangerous to include high-$\ell$ CMB data in the search for
gravitational waves, as the spectral indices may not be constant.
Conclusions about the amount of relic gw in the interval
$2\le\ell\le100$ depend on the contribution of density
perturbations to these multipoles, but the assessment about the
participating $A_s$ and $n_s$ may be wrong, if it is built on the
hypothesis of constant spectral indices in a huge interval of the
observed multipoles. The relatively small number of data points in
the interval $2\le\ell\le100$ brings forth the increased
uncertainty in the evaluation of relic GW, but it seems to be
wiser to live with larger uncertainty (and wait for better data)
than with artificial certainty based on wrong assumptions.

The arguments put forward in \cite{zbg}, as well as the results of
Sec.\ref{run}, hint at possible deviations of the underlying spectra of
primordial perturbations from strict power-laws. We perform a special
investigation of this issue in Sec.\ref{sec4}.

No doubt, in a search for relic GW, one should always check for
the presence of residual systematic effects and alternative
explanations. The lower-$\ell$ multipoles attract attention for a
variety of reasons, including purely instrumental deficiencies,
and they all should be examined (see, for example,
\cite{coldspots}, \cite{quadrupole}, \cite{parity}). The helpful
signatures of gravitational waves are a nonzero B-mode of CMB
polarization \cite{zaldarriaga1997} and a negative $TE$
cross-correlation at lower multipoles \cite{zbg09}.

In Sec.\ref{sec3.1} we make exactly the same assumptions as in
\cite{zbg}, including the fixed best-fit values of the background
parameters. But the new results for the perturbation parameters
$A_s$, $n_s$ and $r$ follow from the CosmoMC technique rather than
from the procedure of Ref.\cite{zbg}. We generalize this search in
Sec.\ref{sec3.2} where the background parameters are not fixed but
are derived together with the perturbation parameters. For this
purpose we use the CosmoMC facility allowing to combine external
data, not affecting the perturbation parameters, with the CMB data
at $2\le\ell\le100$.

\subsection{Fixed background parameters \label{sec3.1}}

To work exactly with the same assumptions as in Ref.\cite{zbg},
we fix the background parameters at their best-fit values of $\Lambda$CDM
cosmology \cite{komatsu}: $\Omega_b h^2=0.02260$, $\Omega_c h^2=0.1123$,
$\Omega_{\Lambda}=0.728$, $\tau=0.087$. The Hubble constant $h=0.704$ is a
derived parameter. The free parameters subject to evaluation by the CosmoMC
code are $\ln (10^{10}A_s)$, $n_s$ and $r$. The spectral indices are related
by $n_t=n_s-1$. The CMB data are used only up to $\ell_{\rm max} = 100$. We call
this set of assumptions the {\it Case~III}.

There exists some awkwardness in our choice of the background
parameters, as these are the parameters that were derived by the
WMAP team \cite{komatsu} under the assumptions of complete absence
of gravitational waves and strictly constant $n_s$ in the entire
interval of participating wavelengths. We know from our previous
experience \cite{zbg} that the background parameters, if changed
not too much, do not greatly affect the results for the
perturbation parameters. Nevertheless, for safety, we explicitly
explore the issue of background parameters, and allow them to
vary, in the more general approach of Sec.\ref{sec3.2}. This will
remove the aforementioned awkwardness in choosing and fixing the
background parameters.

By running the CosmoMC code, we find the following 3d ML values of
the perturbation parameters $r$, $n_s$ and $\ln (10^{10}A_s)$:
\beqa \label{3d} r=0.285,~~n_s=1.052,~~\ln (10^{10}A_s)=3.023,
\eeqa and $n_t=0.052$. We also derived the marginalized 1d results
for these parameters (see also Table \ref{table2}), \beqa
n_s=1.064^{+0.058}_{-0.059},~~\ln
(10^{10}A_s)=2.996^{+0.108}_{-0.112}. \eeqa Following the
convention of \cite{komatsu}, we quote here and below the mean
values of the 1d likelihood functions, and the uncertainties refer
to the $68\%$ confidence intervals. The 1d likelihood functions
for $n_s$ and $r$ are plotted by red lines, also marked by (3), in
the upper and lower panels of Fig. \ref{figure1}, respectively.

The red curve in the lower panel shows a clearly visible broad maximum at
$r=0.2$. Unfortunately, the peak point $r=0.2$ is not strongly separated
from $r=0$, it is nearly (slightly less than) 1$\sigma$ away from $r=0$.
Speaking more accurately, the
$68\%$ and $95\%$ areas under the probability curve are covered,
respectively, by the following  intervals of the parameter $r$:
\beqa
r\in(0,~0.452)~~~{\rm and}~~~r\in(0,~0.843),
\eeqa
so that the point $r=0$ (barely) belongs to the $68\%$ interval.

The 3d ML values, $n_s = 1.052$ and $r=0.285$, obtained with
CosmoMC are smaller than those found in our previous work
\cite{zbg}, $n_s=1.111$ and $R= 0.264$ (which is equivalent to
$r=0.550$), in exactly the same setting. At the same time, the ML
value of $A_s$ is larger than before. Physically, this means that
the new analysis indicates a larger contribution of density
perturbations and a smaller contribution of gravitational waves.
The outcome $r=0.285$ in (\ref{3d}) is almost a factor of 2
smaller than the previous number $r=0.550$. Taken for the face
value, these new evaluations weaken the indications of relic
gravitational waves in the WMAP7 data from approximately a
2$\sigma$ level to approximately a 1$\sigma$ level. This also
worsens the prospects of confident detection of relic GW by the
Planck satellite (more details in Sec.\ref{sec6}). Nevertheless,
it is fair to say that some qualitative indications of blue
spectral indices, and possibly of a large amount of gravitational
waves, have survived.

It is unclear to us why the CosmoMC numerical technique has led to
a factor of 2 grimmer results for $r$ than the previous analysis in
\cite{zbg}. It does not seem likely that this happened because of a
better treatment of noises. The increased noises would probably
lead only to a larger spread of likelihood functions, but this does not seem
to be the case. The difference is more like a systematic shift of ML
points for $r$ by a factor of 2 toward smaller values of $r$. More work
is needed in order to understand the cause of this discrepancy.

\begin{table*}
\caption{Results for $n_s$ and $r$ in Case III and Case IV}
\begin{center}
\label{table2}
\begin{tabular}{|c|c|c|c|c|c|}
  \hline
  \multicolumn{1}{|c|}{  } & \multicolumn{2}{c|}{Maximum likelihood}& \multicolumn{2}{c|}{1-d likelihood}  \\
         \hline
 & $n_s$& $r$& $n_s$ &$r$~($95\%$C.L.)\\
          \hline
 Case III & $1.052$ & $0.285$& $1.064^{+0.058}_{-0.059}$ & $r<0.843$ \\
           \hline
 Case IV & $1.075$ & $0.313$& $1.064^{+0.055}_{-0.053}$ &$r<0.471$\\
  \hline
\end{tabular}
\end{center}
\end{table*}

\subsection{Varied background parameters \label{sec3.2}}

To alleviate the worries about fixed background parameters, we run the
CosmoMC option allowing to find the background parameters together with the
perturbation parameters. The information on the background parameters is
provided by the widely quoted external data (not affecting the perturbation
parameters). We now set free not only the
perturbation parameters $r$, $n_s$, $\ln (10^{10}A_s)$, but
also the background parameters $\Omega_{b}h^2$, $\Omega_{c}h^2$, $\tau$,
$\theta$. Together with the WMAP Collaboration \cite{komatsu}, we include in
the code the external data on $H_0$ \cite{hst}, BAO \cite{hst,bao},
and SNIa \cite{sn}. In addition, we impose a prior on the ``age of the
Universe" $t_0\in(10,~20)$Gyr. The CMB data are used only up to
$\ell_{\rm max} = 100$. We call this set of assumptions the {\it Case IV}.

The analysis has led us to the following marginalized 1d values of the
background parameters:
\beqa
\Omega_b h^2&=&0.02277^{+0.01906}_{-0.01962}, ~~~
\Omega_c h^2=0.1246^{+0.0236}_{-0.0229}, \nonumber\\
\Omega_{\Lambda}&=&0.712^{+0.039}_{-0.030}. ~~~
\tau=0.095^{+0.067}_{-0.080}, \nonumber
\eeqa
They are quite close to the values that we used as fixed parameters
in the {\it Case III}.

\begin{figure*}[t]
\begin{center}
\includegraphics[width = 13cm,height=10cm]{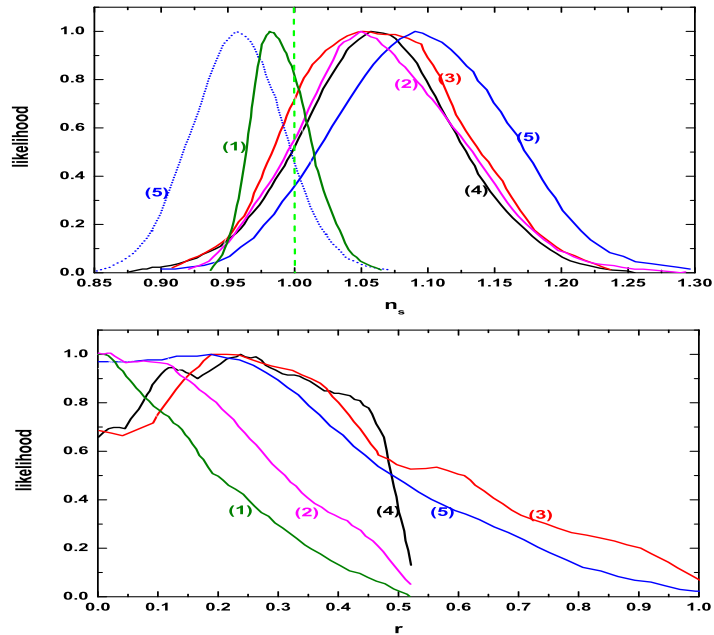}
\end{center}
\caption{One-dimensional likelihood functions for $n_s$ (upper
panel) and $r$ (lower panel). In both panels, the olive curves,
also marked by (1), denote the results for {\it Case I}, the
magenta curves (2) denote the results for {\it Case II}, the red
curves (3) for {\it Case III}, and the black curves (4) for {\it
Case IV}. In both panels, the blue curves, also marked by (5),
denote the results for the {\it Case V}, i.e. for the case with
piecewise power-law spectrum. In the upper panel, the blue solid
line (5) shows the likelihood function for $n_s^{(1)}$, whereas
the blue dotted line (5) shows the likelihood function for
$n_{s}^{(2)}$. In the lower panel, the blue line (5) shows the
likelihood function for $r$.}\label{figure1}
\end{figure*}

As for the perturbation parameters, the 7d ML point was found at
\beqa \label{3d4} r=0.313,~~n_s=1.075,~~\ln (10^{10}A_s)=3.026,
\eeqa and $n_t=0.075$. We also derived the 1d mean results (see
also Table \ref{table2}): \beqa n_s=1.064^{+0.055}_{-0.053},~~\ln
(10^{10}A_s)=3.049^{+0.083}_{-0.085}. \eeqa The 1d likelihood
functions for $n_s$ and $r$ are plotted by black curves, also
marked by (4), in Fig. \ref{figure1}. The distribution for $r$
again, like in the {\it Case III}, shows a maximum at $r=0.2$. The
$68\%$ and $95\%$ intervals are covered, respectively, by \beqa
r\in(0,~0.332)~~~{\rm and}~~~r\in(0,~0.471). \eeqa

The reported numbers, as well as the shapes of red ({\it Case III}) and
black ({\it Case IV}) curves in  Fig. \ref{figure1}, are
pretty close to each other. The more concentrated form of the likelihood
function for $r$ depicted by the black curve, as compared with
the red curve, is probably the result of removal of artificial covariances
between the background and perturbation parameters implicit in the fixed
background parameters assumption of the {\it Case III}. The {\it Case IV}
is superior to the {\it Case III} in that its assumptions are more general.
And, yet, the results for $n_s$ and $r$ turned out to be quite similar. The
spread of distributions for $r$ shown by red and black curves is still
considerable, and the likelihoods include (barely) the point $r=0$ in the
$68\%$ C.L.. Nevertheless, the red and black curves are very different
from the olive curve in Fig. \ref{figure1} ({\it Case I}).

The results of the {\it Case I} are usually interpreted as a
nondetection of GW with a firm upper bound on $r$. In contrast to
the {\it Case I}, the results of data analysis performed along the
lines of {\it Case III}, and especially {\it Case IV}, justify our
previous conclusion, based now on CosmoMC calculations, that there
does exist a hint of presence of relic gravitational waves at the
level of $r \approx 0.2$. Of course, this is a hint, but not a
reliable detection.


\section{Judgments about $n_s$ and $r$ as functions of $\ell_{\rm max}$
\label{sec4}}

The assumption of strictly constant spectral indices is a theoretical
possibility, but not a necessity, and therefore it should be tested by
observations. Here, we show how the judgments about $n_s$ and $r$ depend
on the volume of the utilized CMB data. Concretely, we adopt the assumption
of constant $n_s$ and $n_t$, Eq.(\ref{PsPt}), but we include in the data
analysis only the CMB data up to a certain, varied in steps, multipole
number $\ell_{\rm max}$:  $2\le\ell\le\ell_{\rm max}$. In other words, the
varied $\ell_{\rm max}$ of this Section replaces the fixed largest
$\ell_{\rm max}$ mentioned in Sec.\ref{const}.

It would be much too time consuming to run the CosmoMC for each
step in $\ell_{\rm max}$. Therefore, for the purpose of this
calculation only, we return to the numerical techniques of Ref.
\cite{zbg}. We set free the perturbation parameters ($r$, $n_s$,
$A_s$) and fix the background parameters at the same best-fit WMAP
numbers ($\Omega_b h^2=0.02260$, $\Omega_c h^2=0.1123$,
$\Omega_{\Lambda}=0.728$, $\tau=0.087$) that were used in the {\it
Case III} of Sec.\ref{sec3.1}. The perturbation parameters are
being determined from the $TT$ and $TE$ CMB data sets truncated at
a given $\ell_{\rm max}$. Since the $TT$ uncertainties are
significantly smaller than the $TE$ uncertainties, the major role
belongs to the $TT$ data.

\begin{figure*}[t]
\begin{center}
\includegraphics[width = 18cm,height=8cm]{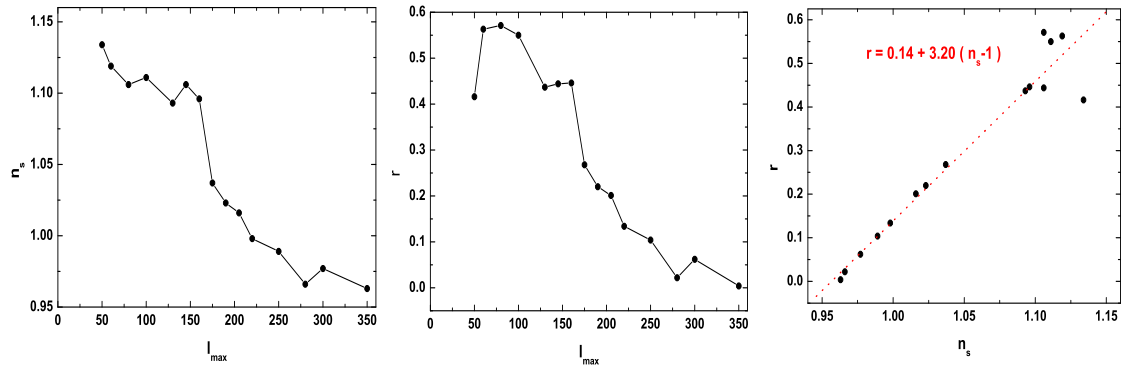}
\end{center}
\caption{The maximum likelihood values of $n_s$ and $r$ as functions of
$\ell_{\rm max}$.}\label{figure2}
\end{figure*}

\begin{figure*}[t]
\begin{center}
\includegraphics[width = 18cm,height=8cm]{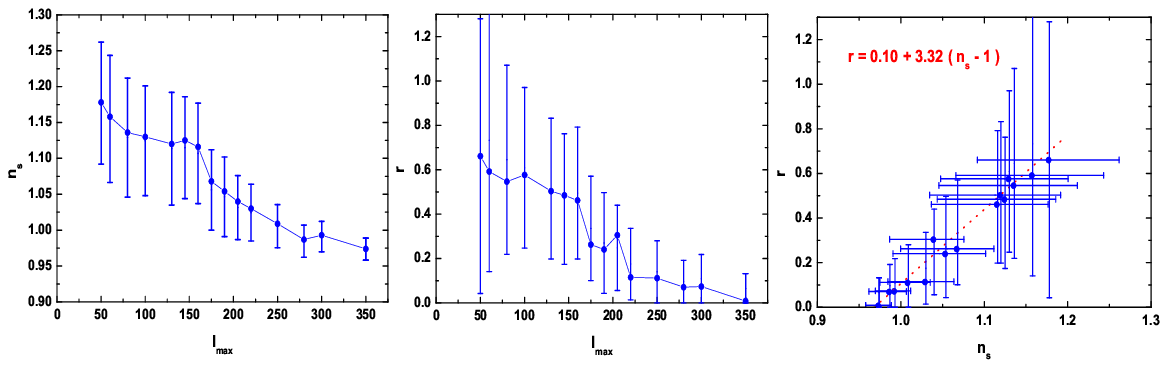}
\end{center}
\caption{The marginalized One-dimensional distributions
(constraints) of $n_s$ and $r$ as functions of $\ell_{\rm max}$.
The vertical bars denote the uncertainties at $68\%$
C.L.}\label{figure3}
\end{figure*}

The 3d ML values of $n_s$ and $r$ as functions of $\ell_{\rm max}$ are shown
in Fig. \ref{figure2}. The right panel exhibits the relation between $n_s$ and
$r$ which arises when the common variable $\ell_{\rm max}$ is excluded. The
corresponding marginalized 1d distributions, showing also the $68\%$
uncertainty bars, are plotted in Fig. \ref{figure3}.

It is clearly seen from Fig. \ref{figure2} and Fig. \ref{figure3} that when
the $\ell_{\rm max}$ is sufficiently large, $\ell_{\rm max} \ge 350$, the
value of $n_s$ approaches $n_s = 0.96$ and $r$ approaches zero. In other
words, under the adopted assumptions we recover the results of {\it Case I}
and the conclusions of the WMAP ``minimal" model \cite{komatsu}.
Specifically, at $\ell_{\rm max} = 350$ we get the 3d ML values $n_s=0.963$
and $r=0.004$.

However, the gradually decreasing $\ell_{\rm max}$ leads to the
gradually increasing $n_s$ and $r$. As soon as $\ell_{\rm max}$
drops to $\ell_{\rm max} \approx 160$, the 3d ML points and 1d
distributions turn to distinctly blue spectra $n_s>1$ and nonzero
$r$. At the point $\ell_{\rm max} = 100$ we return to the 3d ML
result of \cite{zbg}: $n_s=1.111$ and $r=0.550$ (equivalent to $R=
0.264$). The sharp drop of $r$ at $\ell_{\rm max} = 50$ is
probably the consequence of small number of participating data
points and strongly increased uncertainties, as illustrated in the
middle panel of Fig. \ref{figure3}.

As was shown above, within the same assumptions, the CosmoMC
technique returns the smaller numbers, $n_s = 1.052$ and
$r=0.285$, at the point $\ell_{\rm max} = 100$. This was the
subject of discussion in {\it Case III} of Sec.\ref{sec3.1}.
Nevertheless, the general trend toward blue spectra and nonzero
$r$ at lower multipoles remains the same, independently of the
applied numerical code.

It is interesting to note that the judgments about $n_s$ and $r$ that we
arrived at satisfy the linear relations: $r=0.14+3.20(n_s-1)$ for 3d ML
parameters (right panel in Fig. \ref{figure2}) and $r=0.10+3.32(n_s-1)$
for 1d likelihoods (right panel in Fig. \ref{figure3}). The increased
scatter of points at the right ends of these lines is the reflection of
increased uncertainties for small values of $\ell_{\rm max}$.

For every fixed $\ell_{\rm max}$, the uncertainties in $n_s$ and $r$
(the vertical bars around central points in the left and middle panels of
Fig. \ref{figure3}) can be interpreted as the ``degeneracy" between $n_s$
and $r$, in the sense that one and the same set of data can be quite
successfully described by slightly different pairs of $n_s$ and $r$. For example,
at $\ell_{\rm max} = 100$ (see the left panel in Fig. 2 of Ref.\cite{zbg}),
the ``degeneracy" is represented by the more or less elliptical 2d-area, with
the major axis providing a linear relation between $n_s$ and $r$. The movement
along this line illustrates the fact that for a given set of data, and
at least for $\ell_{\rm max} = 100$, a slightly
larger $r$ requires a slightly larger $n_s$, and a slightly smaller $r$
requires a slightly smaller $n_s$. In terms of the left and middle panels
of Fig. \ref{figure3}, this would be the movement along the vertical bars
around the central point at a given $\ell_{\rm max}$.

It is important to note that the inclined lines in the right panels of
Fig. \ref{figure2} and Fig. \ref{figure3} have a different meaning. They are
the result of movement in the horizontal, rather than vertical, direction,
that is, they represent the central values of $n_s$ and $r$ for different sets
of data characterized by different limiting $\ell_{\rm max}$. If the hypothesis
of a strictly constant $n_s$
(and a fixed $r$) were true in the entire interval of accessible multipoles,
we  would expect the inclined lines in the right panels of
Fig. \ref{figure2} and Fig. \ref{figure3} to degenerate to a point,
surrounded by some uncertainties. But this did not happen. We interpret this
fact as a hint of a genuine, even if a very simple, dependence of $n_s$ on
spatial scale.

\begin{table*}
\caption{Results for $n_s^{(1)}$, $n_s^{(2)}$ and $r$ in Case V}
\begin{center}
\label{table3}
\begin{tabular}{|c|c|c|c|c|c|c|c|}
  \hline
  \multicolumn{1}{|c|}{  } & \multicolumn{3}{c|}{Maximum likelihood}& \multicolumn{3}{c|}{1-d likelihood}  \\
         \hline
 & $n_s^{(1)}$& $n_s^{(2)}$ & $r$& $n_s^{(1)}$ &$n_s^{(2)}$& $r$~($95\%$C.L.)\\
          \hline
 Case V & $1.067$ &$0.936$& $0.113$& $1.095^{+0.062}_{-0.061}$ & $0.958^{+0.032}_{-0.032}$ & $r<0.720$ \\
           \hline
\end{tabular}
\end{center}
\end{table*}


\section{Piecewise power-law spectra \label{sec5}}

The findings described in previous Sections support the
proposition that it is unwise to postulate one and the same
power-law spectrum of primordial perturbations for all wavelengths
covered by the existing data. It is risky to do this independently
of the issue of relic gravitational waves. But what is more
important for us, we have shown that if this postulate is imposed,
any signs of GW in CMB disappear, see Sec.\ref{sec4} and {\it Case
I} in Sec.\ref{const}. On the other hand, it is also true that it
is easier to demonstrate the drawbacks of a given hypothesis than
propose a better one. In the absence of a firm guidance, we will
explore the hypothesis which allows the primordial spectrum to
consist of two power-law pieces. This means that we adopt the GW
and DP power spectra in the form of
Eqs.(\ref{break1})-(\ref{break4}). We want to show that this
hypothesis is very much consistent with the indications of GW at
lower multipoles, as found in Sec.\ref{sec3}, and with the
preference for a blue spectrum at longer wavelengths and a red
spectrum at shorter wavelengths, as was discussed in Sec.\ref{run}
and Sec.\ref{sec4}.

To be as close as possible to the already performed calculations,
we use $k_1=0.01$Mpc$^{-1}$ in Eqs.(\ref{break1})-(\ref{break4}).
This choice of $k_1$ corresponds approximately to the multipole
$\ell=100$ \cite{zb}. Certainly, there is no reason for the
piecewise spectrum to be discontinuous at the wave number $k_1$,
and Eqs. (\ref{break1})-(\ref{break4}) take care of this. In
general, the parameters $A_s^{(2)}$ and $A_s^{(1)}$ are not equal
to each other, they are linked by the relation shown in
Eqs.(\ref{break1})-(\ref{break4}). \footnote{In contrast, the
recent paper \cite{norway} explores, possibly due to an oversight,
a strange discontinuous spectrum, where the spectral index $n_s$
is taken as a ``steplike" index, but the parameter $A_s$ is one
and the same in both pieces of the spectrum.}

We again use the CosmoMC sampler. The objective is to build the likelihood
function in the 8-dimensional parameter space, consisting of four background
parameters ($\Omega_bh^2$, $\Omega_{c}h^2$, $\tau$, $\theta$) and four
perturbation parameters ($\ln (10^{10}A_s^{(1)})$, $n_s^{(1)}$,
$n_s^{(2)}$, $r$). The parameter $A_s^{(2)}$ is expressible in terms of
$A_s^{(1)}$, while the spectral indices $n_s$ and $n_t$ in both parts of the
spectrum are related by $n_t = n_s - 1$. The parameters $n_s^{(1)}$ and
$n_s^{(2)}$ are not necessarily different; it is the maximum likelihood
analysis of the data that will tell us their preferred values. Obviously, if
it is postulated that $n_s^{(1)} \equiv n_s^{(2)}$, or if it happens that
$n_s^{(1)} = n_s^{(2)}$ in a particular realization of random trials,
then we simply return to the outcomes of the already considered
{\it Case I} in Sec.\ref{const}. We call the introduced set of assumptions
the {\it Case V}.

We again take into account all the WMAP7 data for $TT$, $TE$, $EE$
and $BB$ correlation functions in the interval
$2\le\ell\le\ell_{\rm max}$, where $\ell_{\rm max}=1200$ for $TT$,
$800$ for $TE$, and $23$ for $EE$ and $BB$. The 8d maximum
likelihood values of the parameters $r$, $n_s^{(1)}$,
$n_{s}^{(2)}$ in the {\it Case V} are found to be \beqa
\label{3dV} r=0.113, ~~n_{s}^{(1)}=1.067,~~n_{s}^{(2)}=0.936.
\eeqa The marginalized 1d likelihood functions for these three
parameters are plotted in Fig. \ref{figure1} by blue curves, also
marked by symbol (5). The 1d results, including the $68\%$
uncertainties, can be summarized as follows (see also Table
\ref{table3}), \beqa \label{1dV}
n_{s}^{(1)}=1.095^{+0.062}_{-0.061},~
n_{s}^{(2)}=0.958^{+0.032}_{-0.032},~ r\in(0,~0.383). \eeqa

Examining the numerical results (\ref{3dV}), (\ref{1dV}) and blue
curves (5) in Fig. \ref{figure1} we can conclude the following.
First, the values for $n_s^{(1)} >1$ and $n_s^{(2)} <1$ confirm
the expectation that the preferred shapes of primordial spectra
are blue at long wavelengths and red at short wavelengths. The
larger (blue) index $n_s^{(1)}$ minus its 1$\sigma$ does not
overlap with the smaller (red) index $n_s^{(2)}$ plus its
1$\sigma$. The value of the (blue) index $n_s^{(1)}$ is about the
same at the values for $n_s$ found in the {\it Case III} and in
the {\it Case IV} for the interval $2\le\ell\le 100$. And the
value of the (red) index $n_s^{(2)}=0.958$ is about the same
(slightly less) than $n_s=0.969$ found in \cite{zbg} for the
interval $101\le\ell\le220$.

The general shape of the likelihood function for $r$ (blue curve
(5) in the lower panel of Fig. \ref{figure1}) is quite similar to
the distributions (3) and (4), plus the expected flattening of the
likelihood function (5) at small values of $r$. The flattening is
expected because there is still a large area in the parameter
space where $n_s^{(1)}$ is equal to $n_s^{(2)}$ and both of them
are less than 1, see the likelihoods (5) in the upper panel of
Fig. \ref{figure1}. As we know from the analysis of the {\it Case
I} all these options return very small values of the parameter
$r$. This is why the probability of small values of $r$ has
increased in comparison with the ``clean" cases, {\it Case III}
and {\it Case IV}, and the likelihood function (5) has flattened
and became consistent with the hypothesis of no gravitational
waves, $r=0$.

The hypothesis of a piecewise spectrum ({\it Case V}) is pretty
much in agreement with WMAP7 data and with all other findings
discussed here. But it also illustrates how careful one should be
in judgments about the absence or presence of gravitational waves
in CMB data. Hopefully, the forthcoming data of better quality
will allow us to make more decisive conclusions.


\begin{figure*}[t]
\begin{center}
\includegraphics[width = 12cm]{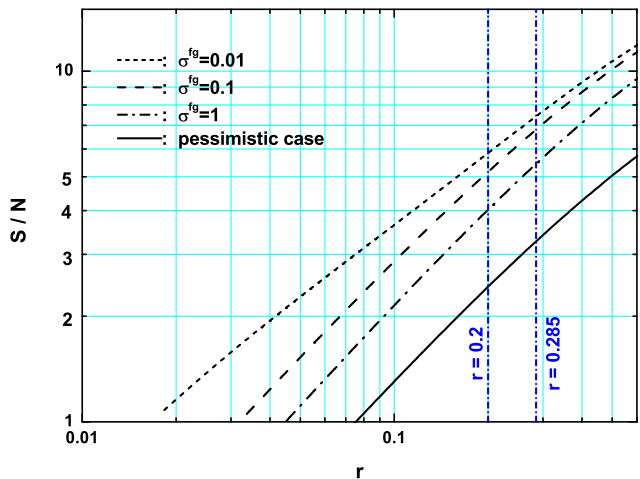}
\end{center}
\caption{The Planck's signal-to-noise ratio $S/N$ for expected GW
signals considered in the text.}\label{figure4}
\end{figure*}

\section{Projections on the Planck mission\label{sec6}}

The CosmoMC calculations have tempered our previous evaluations
\cite{zbg} of relic GW in WMAP7 data. The indications became
weaker, but we do not think they disappeared altogether. The
projections on the possible Planck findings are anticipated to
become worse, but it is important to see what they are now. The GW
signal may have declined in strength, but the good news is that
Planck is definitely expected to operate for 28 months rather than
for 14 months. The longer duration of observations will partially
compensate the suspected weakening of the GW signal.

Following \cite{zbg}, we use the signal-to-noise ratio
\beqa
\frac{S}{N}\equiv \frac{r}{\Delta r},
\eeqa
where the uncertainty $\Delta r$ depends on statistics and various
instrumental and environmental noises. All our input assumptions about
Planck's instrumental noises, number and specification of frequency channels,
foreground models and residual contamination, sky coverage, etc., are exactly
the same as in our previous papers \cite{zbg}. In particular, the ability,
ranging from excellent to none, of removing the foreground contamination is
characterized by the parameter $\sigma^{\rm fg}=0.01, ~0.1, ~1$. We also work
with the pessimistic case, in which $\sigma^{\rm fg}=1$ and the nominal
instrumental noise in the $BB$ information channel at each frequency is
increased by a factor of 4.

The uncertainty $\Delta r$ is evaluated with the help of the
Fisher matrix formalism \cite{fisher}. Since all three
perturbation parameters $r$, $n_s$ and $A_s$ are supposed to be
determined from the same set of Planck's data at $2\le \ell \le
100$, we calculate the $rr$-element of the inverse Fisher matrix,
$\Delta r=\sqrt{(F^{-1})_{rr}}$. All information channels, i.e.
$TT$, $TE$, $EE$ and $BB$ correlation functions, are taken into
account. The data from the frequency channels at $100$, $143$, and
$217$ GHz are supposed to be used, and the observation time of the
Planck mission is taken as $28$ months.

We illustrate in Fig. \ref{figure4} the results of the revised calculation
of $S/N$ as a function of $r$. The two benchmark values of $r$ derived from
the CosmoMC analysis are shown: $r=0.285$, which is the ML value of $r$ found
in the {\it Case III}, and $r=0.2$, which is the 1d peak value of $r$, which
is found in the {\it Case III} and {\it Case IV}. Assuming that $r=0.285$ or
$r=0.2$ are fair representations of the reality, the prospects of discovering
relic gravitational waves with Planck are still encouraging.

If the true value of the parameter $r$ is $r=0.285$, the $S/N$ becomes
$S/N=7.43,~6.81,~5.43$ for $\sigma^{\rm fg}=0.01,~0.1,~1$, respectively.
Even in the pessimistic case, the detection looks quite confident, because
the expected level is $S/N=3.27$. On the other hand, if $r=0.2$ is the true
value of $r$, the $S/N$ diminishes to $S/N=5.87,~5.19,~4.05$ for
$\sigma^{\rm fg}=0.01,~0.1,~1$, respectively. In the pessimistic case, the
prospects of detection drop to $S/N=2.45$.

To conclude, the CosmoMC-revised values of $r$ make the forecasts
for the Planck mission worse than previously evaluated in
\cite{zbg}. It looks like some luck will be needed in the context
of the foreground removal. Nevertheless, even in the pessimistic
scenario, the $S/N$ remains at the interesting level $S/N>2$. It
is also necessary to remember that the search specifically for the
B-mode of CMB polarization by the ground-based and suborbital
experiments (see, for example \cite{bicep}) provides an important
extra avenue for the detection of relic gravitational waves.


~

\section{Conclusions\label{concl}}

We have reanalyzed the WMAP7 data with the help of CosmoMC package
and have shown that the (marginal) indications of relic
gravitational waves are still present. It is vitally important not
to overlook relic GW in the forthcoming data of better quality.
The GW signal is weak and its discovery can be done,
realistically, only by parametric methods. Therefore, a correct
theoretical model and adequate data analysis techniques are
especially crucial. We have stressed the importance of looking for
GW in the lower-$\ell$ interval of multipoles and the dangers of
unwarranted assumptions about density perturbations. The imminent
release of the results of Planck observations will hopefully
confirm our expectations.


\section*{Acknowledgements}
We acknowledge the use of the Legacy Archive for Microwave
Background Data Analysis (LAMBDA) \cite{lambda} and the CosmoMC
package \cite{cosmomc}. Numerical calculations have been done at
the facilities of Niels Bohr Institute and Danish Discovery
Center. We are very grateful to P.Naselsky and J.Kim for
invaluable help in CosmoMC analysis and discussions. W.Z. is
partially supported by Chinese NSF Grants No. 10703005, No.
10775119 and No. 11075141. We thank the anonymous referee for
useful comments.


\end{document}